\documentclass[amsmath, preprintnumbers, 10pt, twocolumn, prl, bibliograpy]{revtex4-1}

\usepackage{graphicx}
\usepackage{subfigure}
\usepackage{color}

\usepackage{physics}

\begin{document}

\title[Nucleation and shape dynamics of model nematic tactoids around adhesive colloids]{Nucleation and shape dynamics of model nematic tactoids around adhesive colloids}

\author{Nicholas B. Ludwig$^1$, Kimberly Weirich$^2$, Eli Alster$^3$, Thomas A. Witten$^4$, Margaret L. Gardel$^4$, Kinjal Dasbiswas$^5$, Suriyanarayanan Vaikuntanathan$^1$}

\email{svaikunt@uchicago.edu}

\affiliation{$1$Department of Chemistry and The James Franck Institute, The University of Chicago, Chicago, IL 60637}
\affiliation{$^2$Pritzker School of Molecular Engineering, The University of Chicago, Chicago, IL 60637}
\affiliation{$^3$ Department of Materials Science and Engineering, Northwestern University, Evanston, IL 60208}
\affiliation{$^4$ Department of Physics and The James Franck Institute, The University of Chicago, Chicago, IL 60637}
\affiliation{$^5$ Department of Physics, University of California, Merced, Merced, CA 95343}

\begin{abstract}
\noindent Recent experiments have shown how nematically-ordered tactoid shaped actin droplets can be reorganized and divided by the action of myosin molecular motors. In this paper, we consider how similar morphological changes can potentially be achieved under equilibrium conditions. Using simulations, both atomistic and continuum, and a phenomenological model, we explore how the nucleation dynamics, shape changes, and the final steady state of a nematic tactoid droplet can be modified by interactions with model adhesive colloids that mimic a myosin motor cluster. Our results provide a prescription for the minimal conditions required to stabilize tactoid reorganization and division in an equilibrium colloidal-nematic setting. 

\end{abstract}

\maketitle
\section{Introduction} \label{sec:tactoidIntro}
Nematic liquid crystals comprise rod-like particles that mutually align along a preferred direction (known as the ``director'') to create a fluid phase with long-range orientational order. The elastic energy cost of deviating from such preferred directions of alignment can be leveraged to sculpt complex free energy landscapes that direct the self-assembly of colloids and nanoparticles~\cite{Smalyukh18}. There is renewed interest in liquid crystals because biological matter, including collections of elongated cells~\cite{Saw2017, Hirst2017}, and the structural components of their cytoskeleton \textendash\, biopolymer filaments such as actin and microtubules~\cite{Sanchez2012, Kumar2018} \textendash\, exhibit nematic order including active matter phases with large-scale flows \cite{Doostmohammadi2018}. Recently, collections of short, rod-like actin filaments have been shown to form nematic droplets with a characteristic elongated tactoid shape~\cite{Weirich17}, that can incorporate the molecular motor myosin to undergo self-organization and shape transformation \cite{Weirich19}. In particular, these nematic droplets of actin can be divided into two equal-sized droplets by clusters of myosin motors that robustly self-organize to the droplet midplane~\cite{Weirich19}. The authors suggest that the droplet deformation can be understood to arise from local realignments of actin filaments by motor activity that cluster the motors and the surrounding actin into an ``aster''-like arrangement \cite{SoareseSilva2011} with actin filaments radiating outwards from the central myosin cluster. Specifically, they model the cluster of motors as an adhesive (``wettable''), spherical colloidal particle that imposes a perpendicular alignment (``anchoring'') on the actin nematic at its surface \cite{DeGennesText}. In this work, we investigate with equilibrium computer simulations the probable intermediate and final minimal energy configurations of such a nematic droplet wetting a colloidal surface, illuminating the assumptions implicit in, and placing strong constraints on, the model presented in Ref.~\onlinecite{Weirich19}.

We construct a minimal model colloid that mimics the geometric constraints imposed by and interactions due to the aster-like object created by the action of the molecular motors. We are interested in probing how our model colloids affect the shapes of nematic droplets within simple equilibrium simulation models that capture the general properties of nematic order without focusing on the detailed molecular features of the actomyosin system, including myosin motor activity. Our choice to neglect motor activity is consistent with the observed suppression of active mechanical forces as individual myosin filaments cluster~\cite{Kumar2018}.  We apply two frameworks \textendash\, the Gay-Berne (or GB) molecular dynamics model~\cite{GayBerne} as well as a coarse-grained continuum phase field model \textendash\, to gain insight into the nucleation and shape changes of tactoids in contact with colloids. 

Here, we report the observation of long-lived dynamic states in which multiple tactoids associate with our model aster-like colloid. Ultimately, these states are found to be unstable within the simple, equilibrium models we explore.  The ground state is a single tactoid associated at its pole with a colloid, resembling prior experimental observations in molecular liquid crystal droplets~\cite{Abbott13,Whitmer13,Abbott14,Rahimi2015}. We identify the feature \textendash\, the formation of a molecularly-thin layer of nematic fluid on the colloid surface \textendash\, that undermines the long-term stability of the two-tactoid state explored in Weirich et al.~\cite{Weirich19}.  We then generalize the phenomenological model from that work to identify possible conditions under which a two-tactoid state may be stable in the absence of active forces. Together, our results provide constraints on the class of detailed equilibrium molecular models that can be used to obtain the stable two-tactoid states that resemble those observed in Ref.~\onlinecite{Weirich19}.

\begin{figure*}
\centering
\includegraphics[scale=0.4]{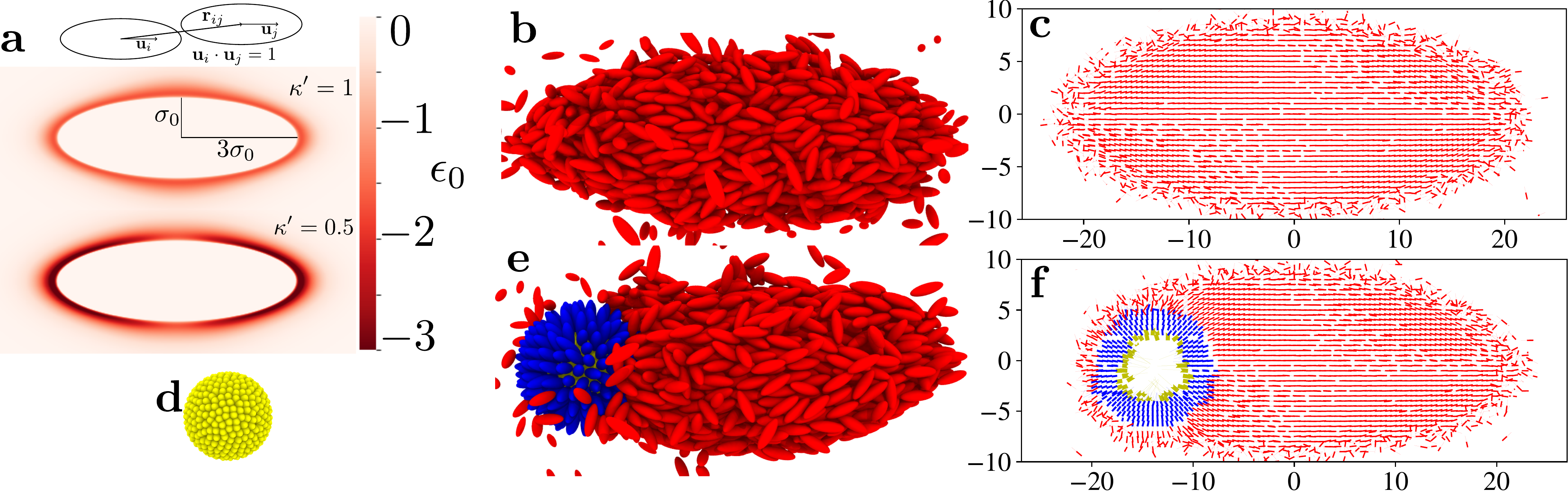}
\caption{Tactoids form from GB rods in molecular dynamics simulations with $T=0.55$, $N=2000$, $L=75\sigma_0$, and can interact favorably with colloids. \textbf{(a)} Schematic for the Gay-Berne model potential for our choices of parameters when rods are parallel. Rods interact based on their distance and orientation, and depend on two aspherical parameters: $\kappa$, the aspect ratio, here set to $\kappa=3$, and $\kappa'$, the ratio of aspherical well depths (see Appendix~\ref{ss:tactoidGBParameterization} for more details). The attractive part of the potentials for two rods with the same orientation and $\kappa'=1$ and $\kappa'=0.5$ are plotted in the second and third portion of this subfigure. The end-end well depth is deeper by a factor of two for $\kappa'=0.5$. Tactoids formed from rods with $\kappa'=1$ are the subject of Section~\ref{ss:tactoidMDFastParameters}, while Section~\ref{ss:tactoidMDSlowSplitting} studies $\kappa'=0.5$. \textbf{(b)} A snapshot from a molecular dynamics trajectory of a tactoid with $\kappa'=1$ from a simulation without a colloid. \textbf{(c)} The director field of the cross-section at the midplane of a tactoid extracted from a molecular dynamics trajectory of a tactoid with $\kappa'=1$ from a simulation without a colloid. \textbf{(d)} A snapshot of the homeotropic colloid used in molecular dynamics simulation. This colloid is composed of $421$ fixed rods with their centers placed on the surface of a sphere with radius $3.5\sigma_0$ and is the colloid we will use in the MD portion of this work. \textbf{(e)} A snapshot from molecular dynamics simulation of a Gay-Berne tactoid with $\kappa'=1$ associated with a homeotropic colloid. In this snapshot, the immobilized colloid is colored yellow, the molecularly-thin splayed nematic layer adsorbed to the colloid is colored blue, and all other rods are colored red. \textbf{(f)} The director field extracted from a molecular dynamics trajectory of a tactoid with $\kappa'=1$ associated with a homeotropic colloid. Same color scheme as (e). All MD snapshots are visualized using Ovito~\cite{Ovito}.}
\label{fig:tactoidSchematicSnapshotDirector}
\end{figure*}

The rest of the paper is organized as follows. We first describe the results of our GB molecular dynamics simulations in Section~\ref{sec:tactoidMDSim}. In Section~\ref{sec:tactoidContinuumSim} we report the results of our continuum phase field simulations, which both complement and extend beyond our MD results. Finally, we outline a phenomenological model that captures some of our simulation observations in Section~\ref{sec:tactoidDiscussion}, discuss the implications of our work, and predict the set of additional features a model would need to allow for stable, divided tactoids.

\section{Colloid-tactoid interactions in a molecular dynamics model} \label{sec:tactoidMDSim}

\subsection{Nematic tactoids associate with aster-like colloids} \label{ss:tactoidMDFastParameters}

Inspired by experimental and theoretical studies on actomyosin clusters~\cite{Weirich19,Husain2017}, in which myosin motors are thought to organize actin filaments into an ``aster''-like (or radial) cluster, we study how an aster-like colloid interacts with tactoid droplets. To simulate rod-like particles in molecular dynamics, we use the GB model~\cite{GayBerne}, implemented in the LAMMPS software package~\cite{Plimpton95,Plimpton09}. GB particles are elongated ellipsoids with an aspect ratio $\kappa$ and with aspherical well depths which are described by a parameter $\kappa'$ of similar form to the aspect ratio; a schematic of the GB potential for parallel rods is shown in Fig.~\ref{fig:tactoidSchematicSnapshotDirector}(a). $\kappa' < 1$ represents stronger end-to-end interactions between the rod-like particles. Our choice to use the GB model was informed by its demonstrated ability to form tactoid droplets with appropriate choice of parameters~\cite{deMiguel97,Fernandez12,Zannoni12} and its simplicity, which makes it a good minimal model. We provide details on our choice of two parameterizations and simulation protocol in the Appendix Section~\ref{ss:tactoidGBParameterization} but briefly, we use an aspect ratio $\kappa=3$, and ratio of aspherical well-depths $\kappa'=1$, which have been well studied in the literature and which ensures that nematic-vapor coexistence occupies a significant region of phase space~\cite{deMiguel97}. Later, in Section~\ref{ss:tactoidMDSlowSplitting} we vary $\kappa'$. Our basic simulation protocol is to start with a vapor of GB rods and to progressively quench the temperature until the system is well within the nematic-vapor coexistence region where tactoids do indeed form, see Fig.~\ref{fig:tactoidSchematicSnapshotDirector}(b,c). There, we show a snapshot from simulation and the director field of the cross-section at the midplane of a tactoid averaged over a trajectory, respectively. Tactoids formed with the parameters discussed now have internal dynamics similar to a liquid.

Our model for a colloid, discussed in more detail in Appendix Section~\ref{ss:tactoidBuildHomeotropicColloid}, is a set of GB particles whose centers of mass are fixed upon the surface of a sphere and are oriented radially, see Fig.~\ref{fig:tactoidSchematicSnapshotDirector}(d). These particles are not allowed to move or rotate during the simulation. This model colloid was constructed to mimic the aster-like arrangement of actin filaments caused by molecular motor action. Throughout this work we report results for a colloid composed of $N_c=421$ fixed particles that share the fluid pair potential and whose centers of mass are located on a sphere of radius $R_c=3.5\sigma_0$ (with $\sigma_0$ the short-axis particle diameter). We tested other colloid sizes and surface densities as well, choosing this size to limit the scale of simulations.

\begin{figure*}
\centering
\includegraphics[scale=0.67]{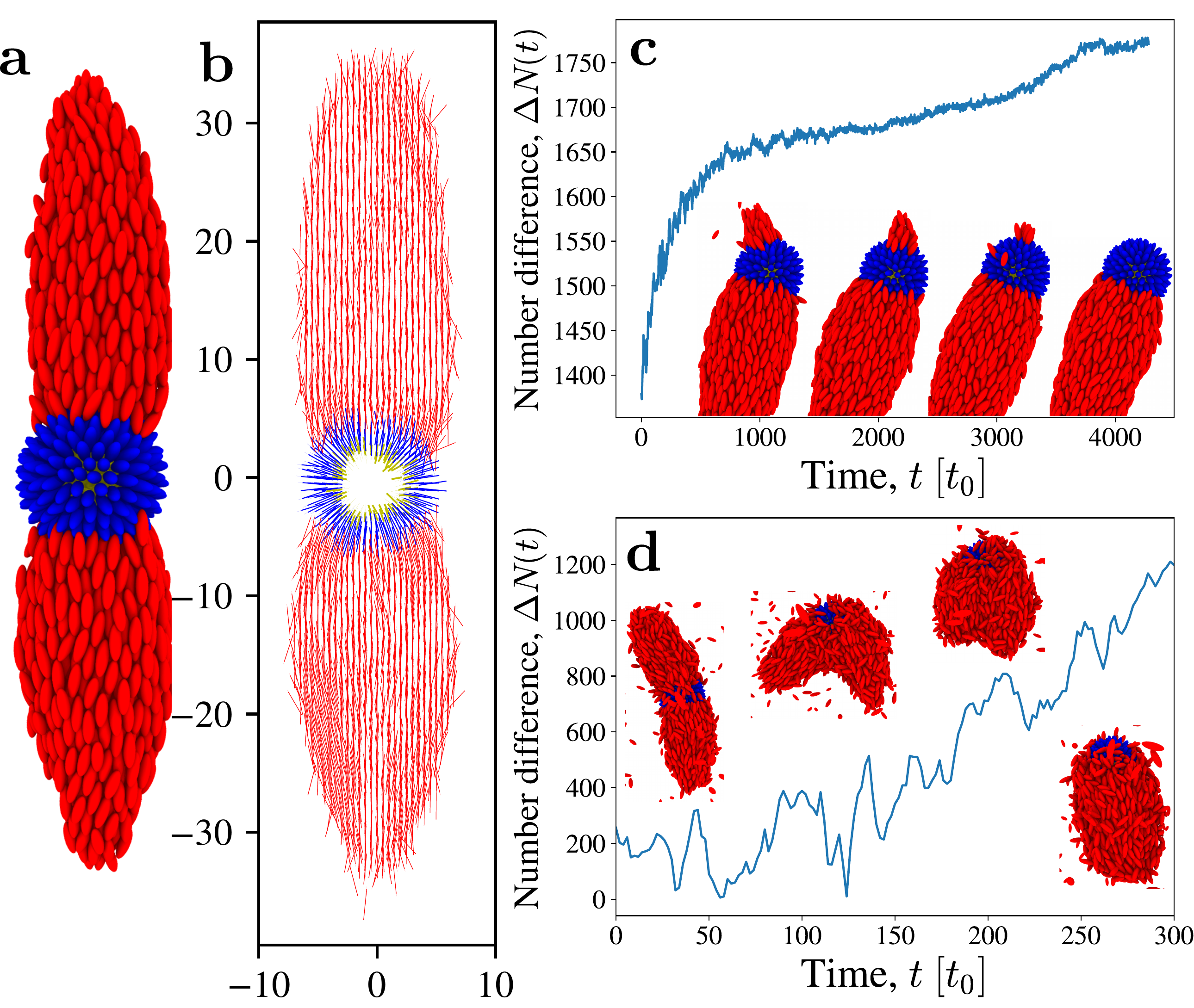}
\caption{In MD, two tactoids may nucleate on the surface of a homeotropic colloid with appropriate choice of parameters, but they are an unstable configuration that gradually evolves into a more stable one.. Here, $N=2000$, $L=96\sigma_0$. \textbf{(a)} A snapshot from a molecular dynamics trajectory of two roughly equal-sized tactoids with $\kappa'=0.5$ associated with a homeotropic colloid. In this snapshot, the immobilized colloid is colored yellow, the thin, splayed nematic monolayer adsorbed to the colloid is colored blue, and all other rods are colored red. \textbf{(b)} The director field of the cross-section at the midplane of two tactoids extracted from a molecular dynamics trajectory of two roughly equal-sized tactoids with $\kappa'=0.5$ associated with a homeotropic colloid. Same color scheme as (a). \textbf{(c)} Time series data showing the difference in particle number between two tactoids for $\kappa'=0.5$. This particular nucleation event resulted in two unequally-sized tactoids at around $t\sim 1000 t_0$. The smaller tactoid slowly loses particles, which are incorporated into the larger tactoid, until it disappears, leaving only the larger tactoid. Snapshots are representative of the decay of smaller tactoid. Same color scheme as (a). \textbf{(d)} Time series data showing the difference in particle number between two tactoids, demonstrating the instability of two tactoids for $\kappa'=1$. The configuration shown in (a) was used as the initial condition, and $\kappa'$ was changed to $1$ at $t=0$. The tactoids equilibrate to their new interaction potential, and then begin to translate on the colloid surface, before combining after around $300 t_0$. Snapshots show a selection of the configurations as the two tactoids combine. Note the rapid timescale relative to (c). Same color scheme as (a). All MD snapshots are visualized using Ovito~\cite{Ovito}.}
\label{fig:tactoidDistributionsTimeSeries}
\end{figure*}

When a vapor of $N=2000$ rods is quenched in the presence of the colloid described above, first a single-molecule-thick layer of radially-oriented rods forms upon the surface of the colloid. In Fig.~\ref{fig:tactoidSchematicSnapshotDirector}(d,e), the rods making up the colloid are colored yellow and those comprising the splayed nematic layer are blue. After the association of the thin splayed nematic layer, a tactoid nucleates upon its surface, leading to a state like shown in Fig.~\ref{fig:tactoidSchematicSnapshotDirector}(e,f), showing a snapshot and the ensemble average director field, respectively. Here, the tactoid has some freedom to translate on the colloid-nematic layer surface, but the rough nature of the surface does not allow unimpeded translation (by ``translation'' we refer to a movement of the interface between the tactoid and the colloid and a corresponding reorientation of the entire tactoid to remain pointing radially outward from the colloid). As has been observed in past studies~\cite{Abbott13,Whitmer13,Abbott14,Rahimi2015}, the tactoid wets the colloid at one end. For an aster-like colloid, such a configuration minimizes the perturbation to the tactoid director field at the cost of reducing the area of the colloid that is wet by the droplet. As suggested by Weirich et al.~\cite{Weirich19}, the area that is wet can be increased by dividing the tactoid droplet into two, at the cost of increasing the surface and elastic energies. In our simple GB model, with $\kappa' = 1$, for tactoid droplets with liquid-like internal dynamics, the adhesive wetting interaction between the colloid and the tactoid is too small to induce such states -- and our ability to increase the wetting interaction strength is limited by the emergence of a molecularly-thin layer. Next, by reducing $\kappa'$, we increase the strength of the attractive interaction between the ends of all GB rods, both fluid and colloid, at constant temperature, in order to: (i), increase the wetting interaction even between the nematic layer and the tactoids, and (ii), slow down the internal dynamics of the tactoid droplets to better examine intermediate, unstable nucleation states.

\subsection{Strong interactions and slow dynamics can lead to transient nucleation of multiple tactoids on a single colloid} \label{ss:tactoidMDSlowSplitting}
To examine intermediate states in the nucleation of tactoids on colloids, and to increase the wetting adhesion strength even in the presence of a molecularly-thin splayed nematic layer, we increase the well depth of the end-end interaction of rods by a factor of two (ie. set $\kappa'=0.5$) while leaving temperature constant (see Appendix~\ref{ss:tactoidGBParameterization} for details and  Fig.~\ref{fig:tactoidSchematicSnapshotDirector}(a) for heatmap plot of potential between parallel rods). Tactoids nucleate similarly to the previous parameter set where $\kappa'=1$, but these have a longer aspect ratio and much slower, more rigid internal dynamics. The tactoids diffuse much more slowly on the surface of the colloid as well. For these parameters, as the initial tactoid grows on the colloid surface, a second tactoid may nucleate as well. This is observed to occur the large majority of the time with $N=2000$ fluid rods in a cubic simulation box with sides $L=96\sigma_0$ and can occasionally lead to two highly symmetric tactoids, see Fig.~\ref{fig:tactoidDistributionsTimeSeries}(a,b) for snapshot and director field. 

Two tactoid states, here, are able to exist on a single colloid not due to a lower free energy, but instead due to the slow internal dynamics and colloid surface translation, relative to the tactoid nucleation and growth timescales. On rare occasions, two fully formed tactoids with $\kappa'=0.5$ were observed to combine by translation on the surface of the colloid. Further, when the tactoids greatly differ in size, the smaller one can be seen to be slowly losing particles until eventually only a single tactoid remains, see Fig.~\ref{fig:tactoidDistributionsTimeSeries}(c) for a particular trajectory particle number time series and snapshots over time. These observations taken together lead us to hypothesize that the ground state for the system in both parameter sets is a single tactoid, and that the two tactoid state is either metastable or is unstable, but with a slow decay to the ground state due to the slow relaxation dynamics of the droplets. 

The instability of the divided drop configuration is more apparent for simulations done with the $\kappa'=1$ parameter set. In Fig.~\ref{fig:tactoidDistributionsTimeSeries}(e), we describe simulations in which we take the divided configuration shown in Fig.~\ref{fig:tactoidDistributionsTimeSeries}(a), generated with the ``slow-dynamics'' $\kappa'=0.5$, and change to the ``faster'' $\kappa'=1$ parameter set. As shown in the time series in Fig.~\ref{fig:tactoidDistributionsTimeSeries}(d), the two tactoids rapidly translate on the surface and combine into one (compare timescales with Fig.~\ref{fig:tactoidDistributionsTimeSeries}(c)). It is clear, then, that for the ``fast'' $\kappa'=1$ parameters, one tactoid is a lower free energy state than two tactoids. 

We note that in order to accurately characterize the stability or metastability of various configurations, it is necessary to compute the free energy of the tactoid-colloid system as a function of order parameters such as the interdroplet angle or the sizes of the various nucleating tactoids. Such free energy calculations are beyond the scope of this work. Rather, in Section.~\ref{sec:tactoidContinuumSim}, we explore using continuum simulations the various potential intermediate and ground states of the tactoid-colloid systems. While the continuum simulations will fail to resolve fine molecular details, they allow us to probe phenomena on longer time and length scales.

\begin{figure*}
\centering
\includegraphics[scale=0.7]{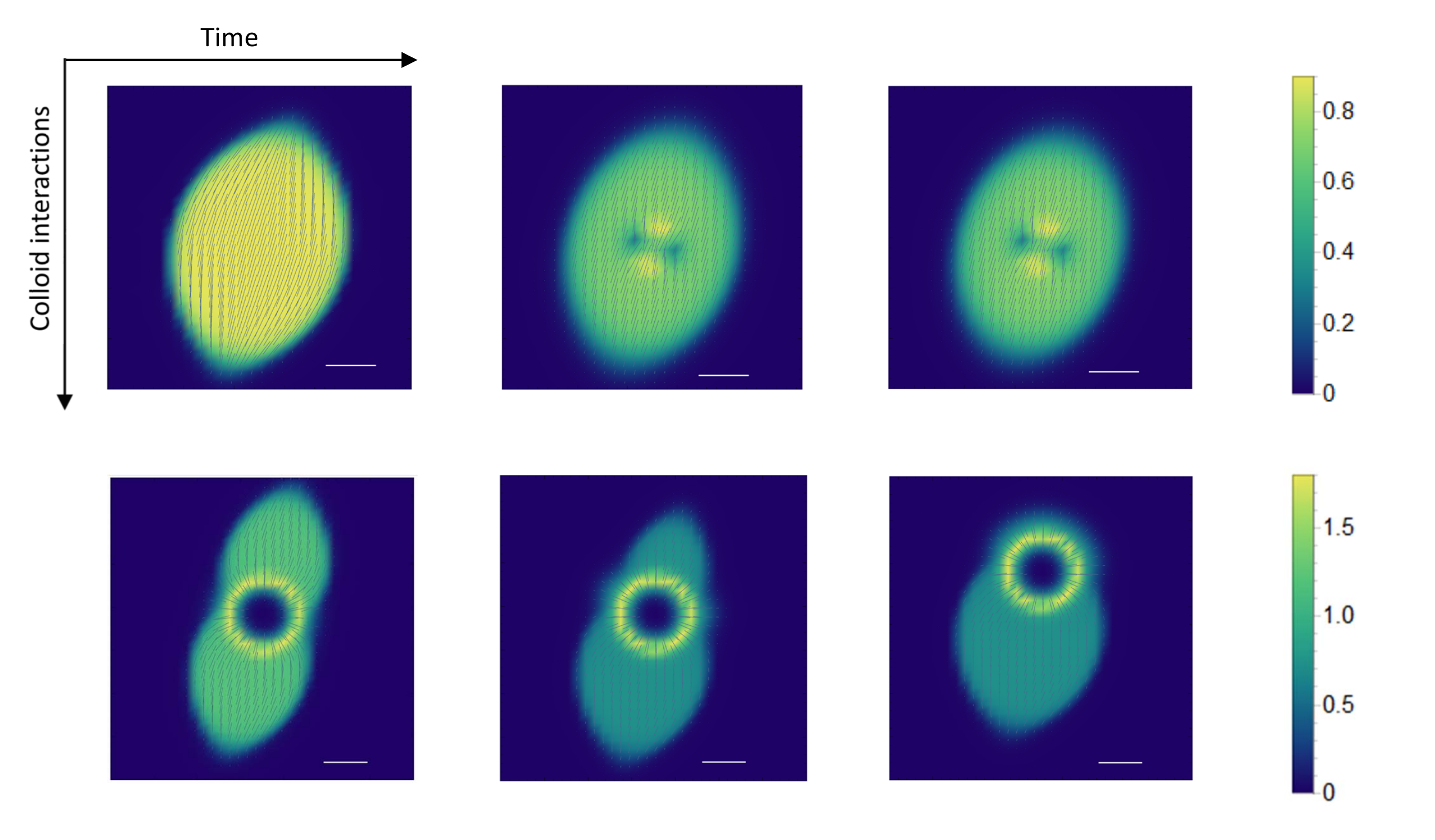}
\caption{Each image shows the nematic density field, ${\bf Q}$ from continuum simulation for the same tactoid droplet, in contact with a colloid that imposes homeotropic (radial) anchoring, at different times during the relaxation of the droplet shape. The tactoid is initialized with the colloid at its center. Colors show the magnitude, $S$, and lines show the director, ${\bf n}$, of the nematic field.  The scale bars in each case correspond to $10$ grid points. The simulation box is $120 \times 120$, only part of which is shown here.  The top panel shows the situation for a smaller colloid (radius, $r_{0}=4$) which has weaker interactions with the nematic. Here, topological defects, seen as regions of depleted nematic order on either side of the colloid, are induced  but the overall tactoid shape is undeformed. This resembles the well-known situation for colloids in bulk nematic. The bottom panel shows the situation for the larger colloid size (radius, $r_{0}=8$) with stronger interactions with the nematic. Here, the tactoid shape can be significantly deformed, resulting in an intermediate divided state. The two divided droplets ultimately coalesce forming a single tactoid with the colloid at its pole. The colloid is not shown directly but corresponds to the region of depleted nematic density at the center of the tactoid where the colloid displaces the fluid.}
\label{fig:tactoidContinuumSimPlots}
\end{figure*}

\section{Tactoids and homeotropic colloids in a continuum phase field simulation} \label{sec:tactoidContinuumSim}

The long length and time scale behavior of the nematic phase, particularly its elastic distortions and defect structures, is traditionally described by the Landau-de Gennes theory~\cite{DeGennesText} which is also used to model nematic-colloid interactions~\cite{Ravnik09}. In contrast with the bulk, nematic droplets resulting from aggregation of rod-like particles have free interfaces that separate the high density nematic from the lower density isotropic or vapor phases. In fact, the alignment of the rods at the droplet interface is a key feature that determines the characteristic tactoid shape~\cite{vdS03}. Thus, a continuum description of the nucleation of tactoids and their shape dynamics should include a density in addition to the nematic order parameter. Here, we adapt a phase field model developed for lyotropic liquid crystals~\cite{Lowen10,Lowen11} to study the interaction between a nematic tactoid droplet and an adhesive colloid. To mimic the binding and arrangement of actin filaments into an aster geometry by a myosin cluster, the model colloid must perform two roles, namely it must interact favorably with the nematic fluid and it must also provide a tendency for the nematic field at its surface to orient radially, also referred to as homeotropic anchoring \cite{Smalyukh18}. We now describe how to construct such a tactoid-colloid model.

Since a nematic tactoid involves bend and splay but not necessarily twist of the director field, we use a 2D description, that is equivalent to looking at a planar section of the full 3D tactoid. The 2D nematic tensor, defined as $ Q_{ij} = S(n_{i} n_{j} -\frac{1}{2} \delta_{ij})$, contains the scalar magnitude, $S= \frac{1}{2}$Tr$\,\mathbf{Q}^2$, and the director, $\mathbf{n}$, or direction of the local nematic order. The model free energy coupling the 2D, nondimensionalized density, $\psi({\bf x})$, and nematic, $Q_{ij}({\bf x})$, order parameter fields is written as \cite{Lowen11},
\begin{equation}
\begin{split}
F &= \int \, d \mathbf{x}  \bigg[ -\frac{v_{2}}{2} \psi^{2} -  \frac{v_{3}}{3} \psi^{3} +  \frac{v_{4}}{4} \psi^{4}  + \frac{B}{2} (\nabla \psi)^{2}  \\
& + \frac{1}{4} Q_{ij}^{2} (C \psi (\psi - 1) + D_{1}) 
+ \frac{S_{1}}{16} Q_{ij}^{2} Q_{kl}^{2} \\
& + \frac{D_{2}}{2} (\partial_{k} Q_{ik})^{2} + B_{3} \partial_{k} \psi \partial_{i} Q_{ik} \bigg]
\label{free_energy}
\end{split}
\end{equation}
which includes the usual Landau free energy terms describing a first order transition in $\psi$, the usual Landau-de Gennes terms describing a nematic transition in ${\bf Q}$, and two terms coupling $\psi$ and ${\bf Q}$. The first of these represents the transition to nematic order at the high density phase. The second with the coefficient $B_{3}$ describes a ``soft'' anchoring or alignment of the nematic director with the external droplet boundary. We work in the ``equal constant'' approximation, where the energy cost of bend and splay are both included in the term with prefactor $D_2$. Note that while this free energy is for a phenomenological phase field model that describes the nematic-isotropic (or nematic-vapor) interface and can nucleate tactoids for suitable parameters, it can also be derived by coarse-graining molecular interactions~\cite{Lowen10,Lowen11}.

The relaxation dynamics of the tactoids are specified by the standard dynamics for the conserved scalar density field (Model B),
\begin{equation}
\partial_t \psi(\textbf{x},t) = \tau_{Q} \nabla^2 \frac{\delta F}{\delta \psi(\textbf{x},t)},
\end{equation}
and the nematic order parameter field (Model A),
\begin{equation}
\partial_t Q_{ij}(\textbf{x},t) = - \tau_{\psi} \frac{\delta F}{\delta Q_{ij}(\textbf{x},t)}.
\end{equation}
where $\tau_Q$ and $\tau_\psi$ are characteristic time scales of the phase dynamics of the nematic and density fields. 
These dynamical equations in suitably nondimensionalized form are solved on a grid with periodic boundary conditions using a pseudospectral scheme with the XMDS software package~\cite{XMDS}. For appropriate choice of parameter values, we see tactoids nucleate from a random initial configuration. See Appendix Section~\ref{ss:tactoidContinuumSimDetails} for details of implementation and table of parameter values used to form tactoids.

To study the effect of a colloid on the droplet structure, we model the spherical colloid by an additional static field, $\phi(x,y) = \frac{1}{2}\left[1 + \tanh(r_{0} - \sqrt{(x - x_{0})^2 + (y - y_{0})^2})/t_{0})\right]$, where $(x_{0}, y_{0})$ and $r_{0}$ specify the center and radius or size of the colloid respectively, and $t_{0}$ is the thickness of the diffuse interface of the colloid as is usual in a phase field model. We incorporate the colloid surface-droplet interaction, anchoring and adhesion, in the free energy. The perpendicular or radial anchoring of the droplet nematic director at the colloid surface can be included in the free energy as the energy cost of deviating away from a preferred value of the nematic tensor at the colloid~\cite{RapiniPapoular69, NobiliDurand92}, which is proportional to $( Q_{ij} -  Q^{0}_{ij})^{2}$. This preferred value of $Q^{0}_{ij} \propto \partial_{i} \phi \partial_{j} \phi$ is normal to the colloid in direction (given by $\partial_{i} \phi$). 
The corresponding free energy term is then defined as $\frac{1}{2} B_4 \partial_{i} \phi Q_{ij} \partial_{j}\phi$. This term is also an effective surface adhesion since the nematic order, ${\bf Q}$, and correspondingly, the density, $\psi$, are enhanced at the colloid surface because of it. An additional term, $W\phi\psi$, leads to exclusion of nematic fluid from the bulk of the colloid, and helps speed up the simulation dynamics.

We initialize the colloid at the center of a tactoid generated in our continuum model, analogous to the situation for actomyosin tactoids reported in Ref.~\onlinecite{Weirich19}, and let the droplet-colloid relax towards its minimal energy state. Fig.~\ref{fig:tactoidContinuumSimPlots} shows the tactoid nematic order parameter at different instants during the course of the simulation for two different sizes of colloid in contact with the same initial tactoid droplet. We find that the relaxation pathway depends significantly upon the relative sizes of the tactoid droplet and the colloid, with a larger perturbation of the droplet shape seen for larger colloids as expected. In fact for larger colloids, the droplet is initially divided into a two-tactoid state as seen in our MD results, Fig.~\ref{fig:tactoidDistributionsTimeSeries}, and as predicted by the model presented in Ref.~\onlinecite{Weirich19}.  For a smaller colloid, the tactoid surface is unperturbed, but a quadrupolar topological defect forms near the colloid surface, see Fig.~\ref{fig:tactoidContinuumSimPlots}, upper panel. This resembles the expected situation for a disk colloid in a 2D bulk nematic~\cite{Silvestre04}. For a fixed tactoid size, increasing the size of the colloid placed into the droplet pushes the defects closer to the nematic-vapor interface, until that interface bows inward toward the defects. The defects can thereby be expelled from the droplet, leading to a state with two tactoid droplets associated with opposite sides of the colloid, see Fig.~\ref{fig:tactoidContinuumSimPlots}, lower panel.  Defect-induced division of a different model nematic droplet was also predicted theoretically in Ref.~\onlinecite{Leoni2017}. At longer times however, the two divided droplets coalesce by diffusing around the colloid and form a single tactoid with the colloid at its pole. This equilibrium state of the system is thus consistent with what is seen in the MD simulations, and is expected for a single tactoid associated with a colloid with strong homeotropic anchoring at its surface.  These observations connect traditional studies of colloid-induced defects in bulk nematics and the deformation and division of tactoid droplets by colloids.

\section{Discussion} \label{sec:tactoidDiscussion}
In Ref.~\onlinecite{Weirich19}, it was proposed that the observed division of actin tactoidal droplets by clusters of myosin motors could be explained by modeling the myosin cluster as a spherical colloid that aligns the actin nematic around it. Specifically, we used the bipolar model for tactoid structure in conjunction with anchoring and adhesion of the actin fluid phase at the myosin cluster interface, to show that a divided droplet state may be energetically favored over a single whole droplet, because it can then increase its area of contact with the colloid. In this paper, we have demonstrated that colloidal interactions may indeed deform nematic droplets, and that even if droplet division is not stable, the ultimate coalescence of two droplets may be slowed down through attractive and radially aligning interactions with a colloid. Specifically, we have reported the observation, within a molecular dynamics and a continuum simulation framework, of dynamic states in which multiple tactoids can be associated with a single colloid. We have shown that these multi-droplet states are unstable within both these models. We find instead that the ground state is a single, whole tactoid associated at its pole with a colloid. In the present section, we generalize the phenomenological model of Weirich et al.~\cite{Weirich19} to explain our results. We will work within the bipolar tactoid model \cite{Taraskin02,vdS03} which has been invoked to explain tactoid shape trends in numerous experiments
\cite{exptAspectRatio1,exptAspectRatio2, exptAspectRatio3, exptAspectRatio4, exptAspectRatio5, exptAspectRatio6, exptAspectRatio7}, as well as simulations \cite{Trukhina2009, Atherton16}, and which our simulated tactoids also resemble.  We then discuss our results in a broader context, and describe a set of minimal extensions beyond the simulation models used here that may allow for the realization of stably divided tactoids. We note, as in Ref.~\onlinecite{Weirich19}, that in the limit that the tactoids are much large than the colloids, the driving forces for tactoid recombination overwhelm any driving forces for tactoid division, since the latter simply depends on colloid size. Hence, in the limit of thermodynamically large tactoid drops, we anticipate that the stable ground state is simply corresponds to single undivided droplet. 

\begin{figure}[t]
\centering
\includegraphics[scale=0.6]{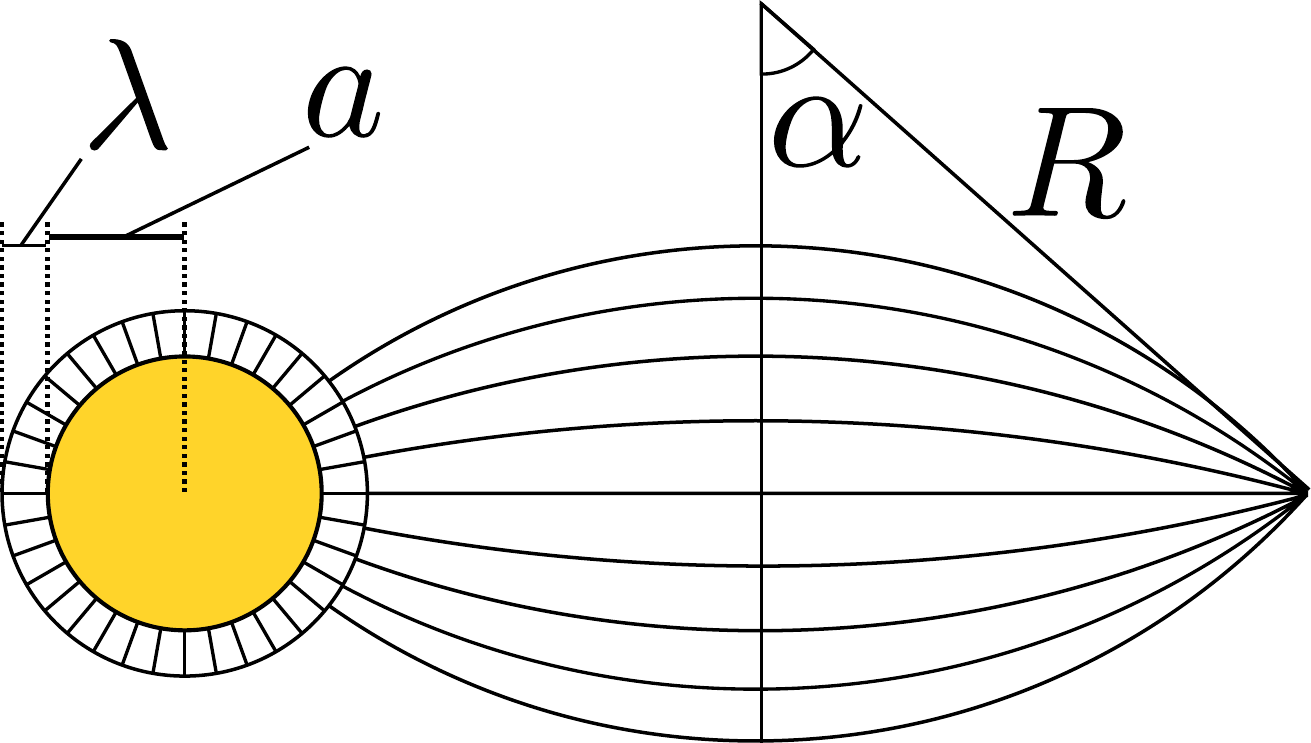}
\caption{A schematic of a tactoid with size and shape parameters: the radius, $R$, and the tip angle $\alpha$, as defined in Ref.~\onlinecite{Taraskin02}. This tactoid is associated at its pole with a colloid of radius $a$ that imposes homeotropic (radial) alignment on the fluid it is in contact with. The colloid surface may be covered by a thin splayed nematic layer of thickness $\lambda$ of the order of the rod length.}
\label{fig:tactoidTheorySchematic}
\end{figure}

\begin{figure}
\centering
\includegraphics[scale=0.48]{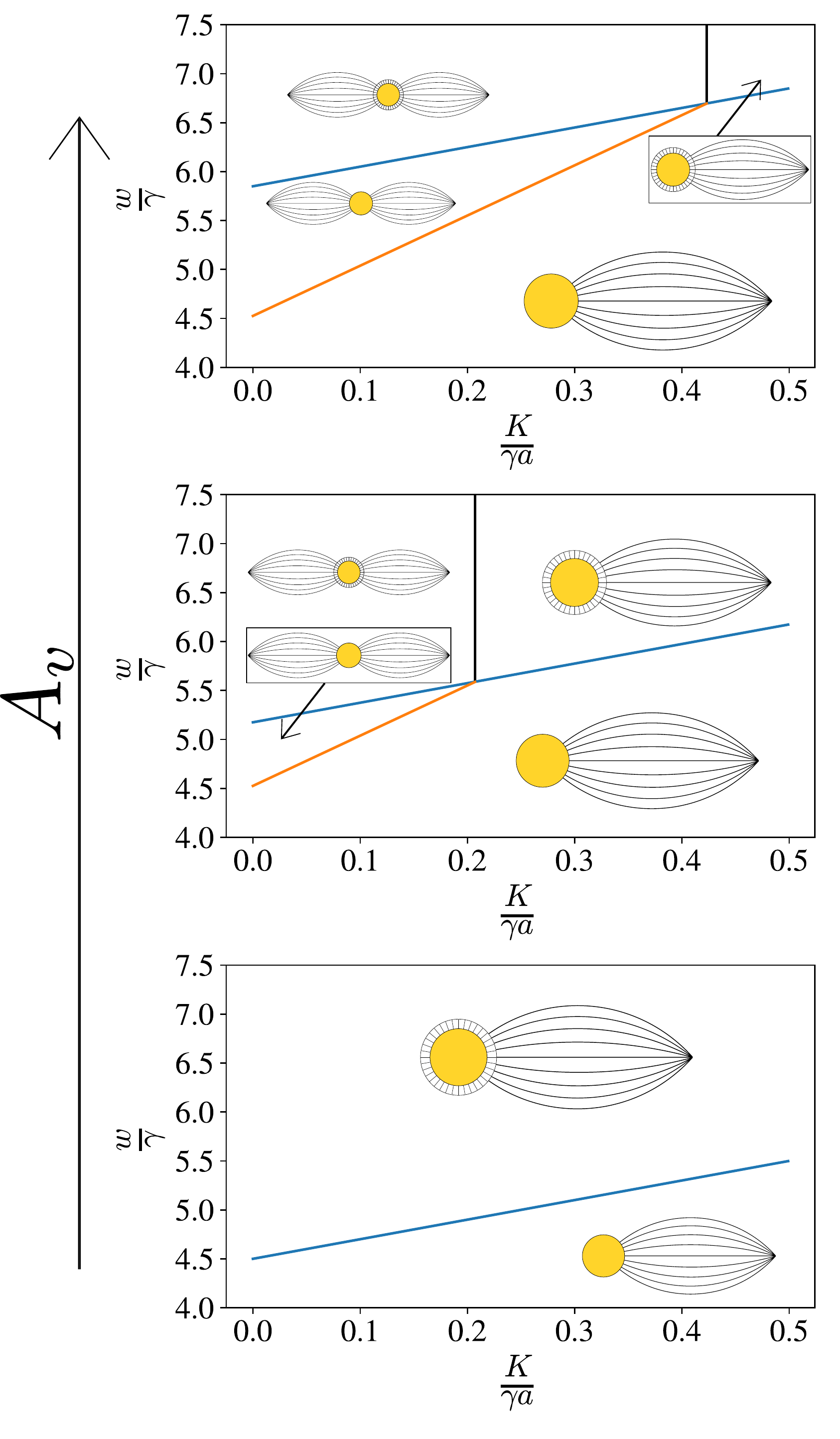}
\caption{Set of phase diagrams for fixed $\lambda=0.5 a$ and total tactoid volume $V_t=2\times 4\pi a^3/3$ with increasing $A_v=1.0 ,\, 1.3,\, 1.6$. The blue line in each plot divides the region where a molecularly-thin layer forms (above the blue line, where $F_{l}<0$ or $w>\epsilon$) from the region where no molecularly-thin layer forms (below, where $F_{l}>0$ or $w<\epsilon$). Above the blue line, the formation of the molecularly-thin layer replaces the wetting parameter, $w$, with a weaker ``effective wetting'' of $\epsilon$, since, upon tactoid association, the area fraction covered by the tactoid, $f_t$ is subtracted from the total area fraction, $f_{l}$ of the molecularly-thin layer in the energy penalty term, ie. $f_{l} \to f_{l} - f_t$, see discussion in main text or Appendix Section~\ref{ss:tactoidWettingLayerFEDerivation}. Thus, the possible existence of the thin nematic layer provides an upper bound on $w$. For a colloid to support (meta)stable divided tactoids, the nematic-vapor interfacial anchoring energy parameter, $A_v$, must be above a certain value. Here, multiple tactoids are a local free energy minimum between the orange and blue lines, for which there is no wetting layer, or above the blue line and to the left of the black line, for which multiple tactoids associate with the wetting layer. However, when $A_v$ is small enough, there is no ``divided'' region, as seen in the phase diagram for $A_v=1.0$.}
\label{fig:tactoidTheoryPhaseDiag}
\end{figure}

A feature not included in the model droplet-colloid free energy in Weirich et al.~\cite{Weirich19} was the possible formation of a molecularly-thin splayed nematic layer on the surface of the colloid, which was a recurring feature observed in our GB and continuum simulations. Instead, it was assumed there the orientation of the director field  was strongly anchored parallel to the droplet nematic-vapor interface, ie. strictly  planar anchoring as in the bipolar tactoid model \cite{vdS03, Taraskin02}. However, such a nematic layer is incompatible with the simultaneous strong planar anchoring at the droplet-vapor interface and strong homeotropic anchoring at the droplet-colloid interface. Relaxing the strong nematic-vapor anchoring constraint allows us to constrain the conditions under which two or more droplets can stably associate with the colloid without coalescing. Thus, motivated to more closely describe the conditions of our present simulation models, we relax this strong nematic-vapor anchoring constraint, by adding an anchoring energy cost, per unit area, at the nematic-vapor interface to the overall free energy $\sim \gamma A_v \left(\textbf{v}\cdot\textbf{n}\right)^2$, with $\gamma$ the surface tension, $A_v$ the dimensionless nematic-vapor anchoring coefficient, $\textbf{v}$ the interface normal and $\textbf{n}$ the director~\cite{RapiniPapoular69,vdS03}. We can then write down the free energy cost of a molecularly-thin splayed nematic layer adhered to the surface of the colloid. Specifically, we consider a layer of radially-directed nematic of thickness $\lambda$ covering a dimensionless fraction $f_{l}$ of the surface area of a colloid of radius $a$, see schematic in Fig.~\ref{fig:tactoidTheorySchematic}. The free energy can be written as
\begin{equation}
\begin{split}
F_{l} &= \left[-w + \frac{4K}{a}\frac{\lambda}{a} + \gamma\left(1 + A_v\right)\left(1+\frac{\lambda}{a}\right)^2\right] f_{l} \, 4\pi a^2 \\
&= \left[-w + \epsilon\right] f_{l} \, 4\pi a^2,
\end{split} \label{eq:tactoidsWettingLayerF}
\end{equation}
with $w$ the strength of the adhesive colloid-nematic wetting interaction, and $K$ the Frank elastic constant in the one-constant approximation (see Appendix~\ref{ss:tactoidWettingLayerFEDerivation} for derivation). In the second line, we have grouped the energy penalties associated with the splayed nematic layer into a single term, explicitly, $\epsilon  \equiv \frac{4K}{a}\frac{\lambda}{a} + \gamma\left(1 + A_v\right)\left(1+\frac{\lambda}{a}\right)^2.$ Thus, a thin splayed nematic layer is favored to form when the adhesive wetting interaction strength between the colloid and the nematic becomes larger than the penalties associated with the stressed nematic state (ie. splay elastic energy, surface energy, and anchoring energy): $w > \epsilon$.

It follows that there are two possibilities we need to consider: the association of one or more tactoids with the colloid in the absence ($F_{l} > 0$), or the presence ($F_{l} < 0$) of a thin splayed nematic layer. The phase diagrams plotted in Fig.~\ref{fig:tactoidTheoryPhaseDiag}, show how the anchoring parameter $A_v$ can be used to change the boundary between these two possible regimes. The first case resembles the model described in Ref.~\onlinecite{Weirich19}, but with an upper bound, $w<\epsilon$, and corresponds to the region below the blue line in each phase diagram. In that model, multi-tactoid states, which increase the colloid-nematic wetting area, can be stabilized when $w$ is larger than a critical value dependent on the elastic and surface energy penalties, $w > w^*$; this regime appears as the region above the orange line in the top two phase diagrams. However, the addition of the upper bound $w<\epsilon$ restricts the size of the stable two-tactoid region. When $w^*>\epsilon$, as is the case for the bottom phase diagram in Fig.~\ref{fig:tactoidTheoryPhaseDiag}, there is no stably-divided region, which may explain why (meta)stable two-tactoid states are not realized in the present simulations. In the second case, when $w>\epsilon$ (above the blue line), a thin nematic layer will form, and tactoid(s) will associate with that layer on the surface of the colloid. There is a free energetic benefit for such an association, but it is not determined by $w$. Instead, the dimensionless area fraction of the thin nematic layer covered by the tactoid, $f_t$, is subtracted from the total nematic layer area fraction in the energy penalty term, leading to a net energy benefit of $-\epsilon f_t \, 4\pi a^2$ upon association. Thus, the possibility of a thin splayed nematic layer sets an upper bound on the ``effective adhesive wetting'' felt by the colloid, potentially eliminating the two-tactoid stable state. Though the nematic layer appears to be a limiting factor in the minimal GB model we study here, we predict that a model which affords control over $A_v$ will allow elimination of the thin nematic layer and access to the stably divided regime. One simple modification to the present GB model is the addition of another particle type that penalizes nematic-vapor interfaces with rods oriented normal to the interface. Such a model could be adapted from work by Moreno-Razo et al.~\cite{dePablo12}.

In summary, the work in this paper shows how colloidal interactions can be used to modify the shape and dynamics of associating liquid crystal tactoid droplets. Our motivation for these studies was the experimentally observed reorganization of of actin tactoidal droplets by clusters of myosin motors in Ref.~\onlinecite{Weirich19}. Our work identifies possible (effectively) equilibrium mechanisms for this observed reorganization and places strong molecular constraints on the model presented in Ref.~\onlinecite{Weirich19}. While active forces resulting from myosin sliding actin filaments may be crucial for the observations in Ref.~\onlinecite{Weirich19}, our studies suggest that engineering suitable liquid crystal-colloidal interactions may also result in nematic droplet deformation.

Both biopolymeric and molecular liquid crystals are candidate materials for future experimental investigations of the deformation of nematic droplets by colloids. Passive beads, as opposed to clusters of motors, can be functionalized to bind actin filaments in aligned orientation \cite{Stachowiak2012}, which under suitable conditions can nucleate tactoids~\cite{Weirich17}. Incorporating strongly homeotropic colloids into molecular liquid crystal droplets may also show the formation of nematic defects in the droplets, and their ultimate expulsion leading to deformation of the droplet interface. However such colloidal inclusions need to have strong affinity for the host liquid crystal in order to be incorporated into the bulk of the droplet, while simultaneously ensuring that the interfacial tension of such liquid crystals is low enough to allow for such deformation. While using the elastic distortion of liquid crystal solvents is a standard route to colloidal self-assembly \cite{Luo2018}, we point out here the possibility of changing droplet shape and nucleation dynamics using colloids. The resulting tunability of droplet morphology may have possible applications in interfacial materials\cite{Stebe2018}.

\begin{acknowledgments}
    This work was primarily supported by NSF DMR-MRSEC 1420709. SV also acknowledges support from the Sloan Foundation and startup funds from the University of Chicago.
\end{acknowledgments}

\section{Appendix}

\subsection{Gay-Berne parameterization and simulation protocol} \label{ss:tactoidGBParameterization}
We study tactoids using molecular dynamics for two parameter sets. Readers are referred to the literature for an introduction of the details of the Gay-Berne model~\cite{Plimpton09}. In both parameter sets, we study uniaxial rods of aspect ratio $\kappa=\sigma_{ee}/\sigma_{ss}=3$, with $\sigma_{ee}$ the particle length and $\sigma_{ss}$ the width, in units of the fundamental length scale $\sigma_0$. The exponent parameters are set to the original parameterization used by Gay and Berne~\cite{GayBerne}, $\mu=2$ and $\nu=1$, which remains a common choice in the literature. The other anisotropic parameter in the uniaxial GB model is the ratio of anisotropic well depths $\kappa'=\epsilon_{ss}/\epsilon_{ee}$, with $\epsilon_{ss}$ the well depth for rods interacting side-to-side and $\epsilon_{ee}$ the well depth for rods interacting end-to-end. We choose $\kappa' = 1$ and $\kappa' = 0.5$ (for which $\epsilon_{ee}=2\epsilon_0$ with $\epsilon_0$ the fundamental energy scale) due to the easier access afforded to the nematic-vapor coexistence portion of the phase diagram for $\kappa'\leq 1$~\cite{Fernandez12}. See Fig.~\ref{fig:tactoidSchematicSnapshotDirector}(a) for a visualization of the GB potential for parallel rods with our choices of parameters. The simulation time step was set to $0.002t_0$, with $t_0$ the natural time scale.

\begin{figure}[h]
\centering
\includegraphics[scale=0.57]{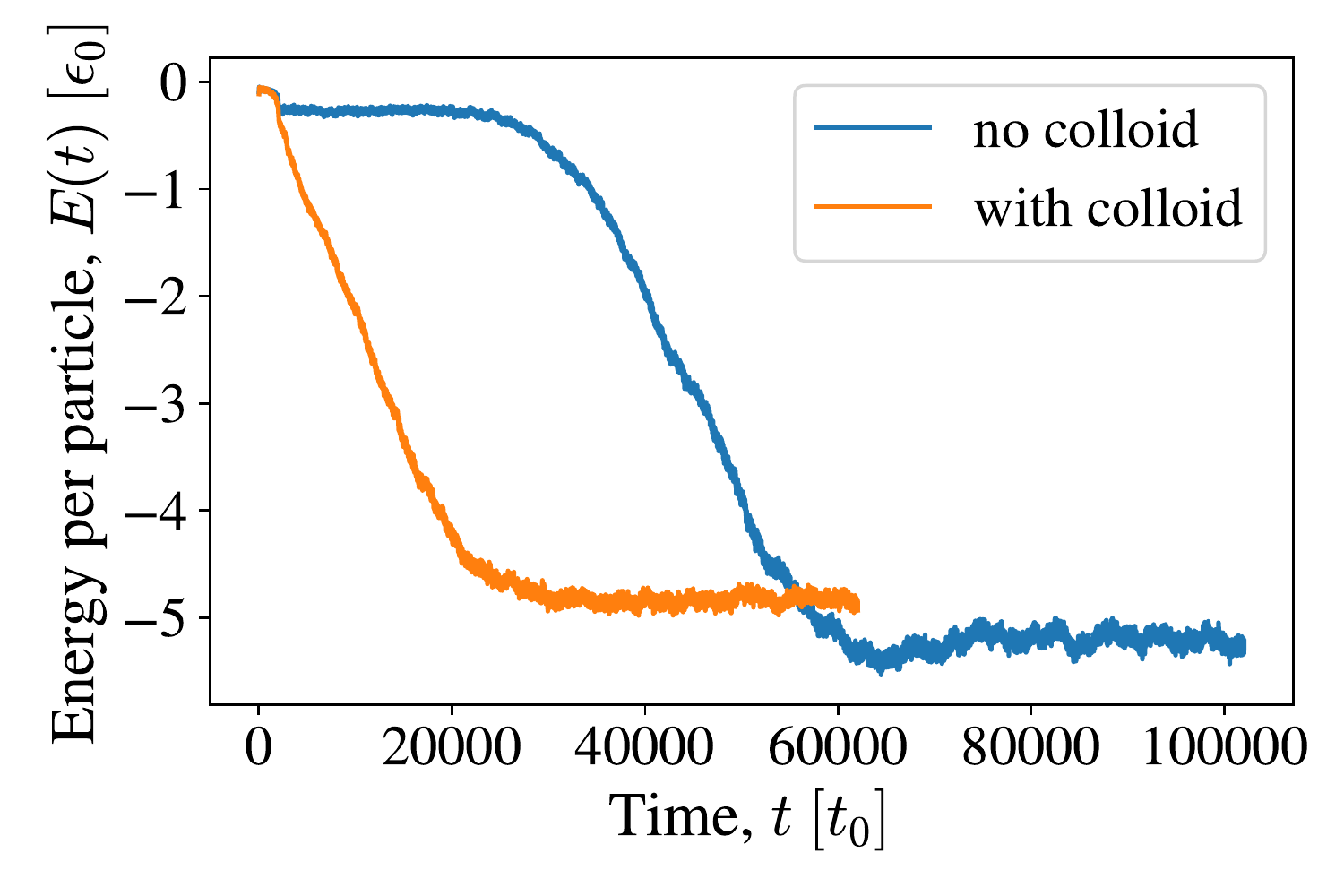}
\caption{Energy time series for $\kappa'=1$, $N=2000$, $L=75\sigma_0$ trajectories without (blue) and with (orange) a colloid demonstrate that equilibrium is reached within a reasonable simulation time. Initial condition is a simple cubic lattice of rods that is vaporized to a gas at $T=2.55$. The temperature is then slowly quenched over time (over the first $1000 t_0$) until reaching its final value of $T=0.55$.}
\label{fig:tactoidKP1WSuTimeSeries}
\end{figure}

Using the parameters mentioned above with $\kappa'=1$, and simulation periodic cube with sides of length $L=75\sigma_0$, a single tactoid quickly forms from a vapor quenched to $T=0.55$, see Fig.~\ref{fig:tactoidSchematicSnapshotDirector}(b). Specifically, the initial condition is an optional colloid and a simple cubic lattice of fluid rods which are vaporized to temperature $T=2.55$ for $10^5$ to $5\times 10^5$ time steps and then quenched in steps of $\Delta T=-0.2$ every $10^5$ timesteps until reaching $T=0.55$, see Fig.~\ref{fig:tactoidKP1WSuTimeSeries}. The director field, which measures the average local orientation of rods in the droplet is extracted, as detailed in the Appendix, and is plotted in Fig.~\ref{fig:tactoidSchematicSnapshotDirector}(c). The tactoid displays liquid ordering, with particles diffusing throughout the droplet. In contrast, the parameter set with $\kappa'=0.5$ leads to tactoids which are much more rigidly ordered and have significantly slower diffusion of particles throughout the droplets, as well as much lower vapor pressure.

\subsection{Implementing a homeotropic colloid in molecular dynamics} \label{ss:tactoidBuildHomeotropicColloid}
We are interested in how a colloid with homeotropic boundary conditions interacts with tactoid droplets. The colloid must perform two roles, namely inducing wetting, and maintaining the boundary condition, or anchoring. We initially used a sphere with a strong wetting interaction with the fluid rods. This approach has met with success in a past study when applied in a bulk nematic, as rod packing led to homeotropic anchoring~\cite{Allen18}. We found that a spherical colloid indeed interacted with a tactoid, but with very weak anchoring regardless of the wetting strength. Further, it was clear that multiple tactoids would not be able to nucleate on the surface of the colloid, as the first tactoid to nucleate was free to translate on the colloid surface and absorb any other nuclei that might form. To improve the strength of the anchoring as well as to add a barrier to translation on the surface, we turned to another model for the colloid. In this model, we place a fixed set of GB particles upon the surface of a sphere and orient them so that they point radially outward, see Fig.~\ref{fig:tactoidSchematicSnapshotDirector}(d). We outline the details of the procedure to create the colloid below in Appendix Section~\ref{ss:tactoidSpacingPointsOnSphere}. Throughout this work we use a colloid composed of $421$ fixed particles whose centers of mass are located on a sphere of radius $3.5\sigma_0$ unless otherwise noted.

\subsection{Finding the near-optimal spacing of points on a sphere} \label{ss:tactoidSpacingPointsOnSphere}
We implement a homeotropic colloid by constructing a fixed set of GB rods with center of masses upon the surface of a sphere and which are oriented radially, see Fig.~\ref{fig:tactoidSchematicSnapshotDirector}(c). We first find the positions of a set of such rods by imagining them as points which we would like to place \textendash\, separated from each other as much as possible \textendash\, on the surface of a sphere. We initialize a given number of points forming the colloid, $N_s$, randomly placed upon the surface of a sphere of radius $R_s$. We estimate the optimal spacing of particles on the surface of the sphere using a well-known iterative approach where we imagine each particle to be an ion interacting with each other ``ion'' according to a repulsive Coulomb $1/r$ interaction. We take Monte Carlo steps, iteratively selecting a particle at random as well as a random move on the surface of the sphere, and then evaluating the probability of accepting the move using the standard Metropolis rule:
\begin{equation}
p_{acc}(\Delta E)=\min\left[1,\,\exp\left(-\beta \Delta E\right)\right],
\end{equation}
with $\beta$ the effective inverse thermal energy, a parameter that increases through the iterative procedure, and $\Delta E$ is the change in energy due to the proposed move, explicitly, for a move of particle $i$,
\begin{equation}
\Delta E = \sum_{j\neq i}^{N_s} 1/r_{ij}^{(proposed)} - 1/r_{ij}^{(old)},
\end{equation}
with $r_{ij} = |\textbf{r}_j-\textbf{r}_i|$. After a sufficient number of Monte Carlo steps, we are left with a good estimate of the configuration of particles that maximizes the space between each particle. We take this set of positions as the positions of the rods which make up the colloid, and for each particle, orient it such that it points radially outward. The result of this procedure for a choice of $N_s=421$ and $R_s=3.5$ is shown in Fig.~\ref{fig:tactoidSchematicSnapshotDirector}(d). To maintain its rigid form throughout the simulation, we simply neglect to time integrate the rods which make up the colloid. Particles in the fluid will interact with the colloid, but the colloid will not translate or change shape as a result.

\subsection{Extraction of director field} \label{ss:tactoidExtractDirectorField}
We extract the director field from MD simulations to gain insight into the perturbation that a colloid induces in the director field. From a time series of trajectory frames, we first exclude any particles that are within the fixed colloid (assuming the particular trajectory has a colloid). Next, we exclude any gas particles by using the following procedure. We compute the adjacency matrix for the system, whose matrix elements are defined as
\begin{equation}
A_{ij}=\exp\left(-\frac{r_{ij}^2}{2\sigma^2}\right),
\end{equation}
with $r_{ij} = |\textbf{r}_j-\textbf{r}_i|$ and $\sigma$ a parameter chosen to make sums over columns $d_i = \sum_j A_{ij}$ large when particle $i$ is in the tactoid droplet and small when in the gas phase (we use $\sigma=0.5$). This is a similar metric to coordination number. We cluster $(d_i+c)^{-1}$ values about two means using the standard $k$-means clustering algorithm, with $c$ a small constant (here, $0.01$). Clustering with $\sim 1/d_i$ was found to give much better results than directly clustering $d_i$. The particles in the cluster corresponding to small $d_i$ values are in the gas phase, and are excluded. Having removed the gas and (optional) colloid particles, we compute the center of mass of the remaining particles, which make up the tactoid droplet. We set the origin to be the center of mass, compute the inertia tensor, and rotate to the inertia tensor eigenvector basis as the natural basis of the droplet. We divide space into discrete bins, and within these bins compute our order parameter of interest: local density or local director field.

The director field is computed within a bin by first computing the nematic tensor order parameter $\textbf{Q}(\textbf{r})$, specifically,
\begin{equation}
Q_{ij}(\textbf{r})=\frac{1}{2N_{b}(\textbf{r})}\sum_k^{N_{b}(\textbf{r})} 3 u_i^{(k)}(\textbf{r}) u_j^{(k)}(\textbf{r}) - \delta_{ij},
\end{equation}
where the sum runs over all particles within a bin, $N_{b}(\textbf{r})$, and $u_i^{(k)}(\textbf{r})$ is the $i$-component of particle $k$'s orientation vector. The director \textbf{n}(\textbf{r}) is the eigenvector corresponding to the largest eigenvalue of $\textbf{Q}(\textbf{r})$, and the largest eigenvalue is the magnitude of the ordering, $S(\textbf{r})$.

Fig.~\ref{fig:tactoidSchematicSnapshotDirector}(b) shows a cross-section of the director field of a tactoid with $N=2000$ particles in a simulation box of size $L=75\sigma_0$. Qualitatively, it is clear that the tactoid is not homogeneous, as the director field tends to mimic the curvature of the nematic-vapor interface. However, the tactoid is also not purely bipolar (compare with the schematic in Fig.~\ref{fig:tactoidTheoryPhaseDiag}(a)). Fig.~\ref{fig:tactoidSchematicSnapshotDirector}(d) shows a cross-section of the director field of a $N=2000$ tactoid in a box of size $L=75\sigma_0 $ associated with a colloid at its lower tip. It can be seen that the colloid templates homeotropic order within a local region, and that even beyond that local region, the director field is perturbed relative to the tactoid without an associated colloid.

\subsection{Details of continuum simulations} \label{ss:tactoidContinuumSimDetails}
\begin{table}[h]
	\centering
	\label{table1}
	\begin{tabular}{|c|c|}
		 \hline
	
		{\bf Parameter} & 	{\bf Value} \\
		
		\hline 
		
	$v_2$  & $0.5$	 \\
    $v_3$  & $0.5$	 \\	
    $v_4$  & $2.0$  \\
    $B$  & $0.5$ \\
    $C$ & $5.0$ \\
    $D_{1}$ & $0.01$\\
    $S_{1}$ & $2.0$ \\
    $D_{2}$ & $1.0$ \\
    $B_{3}$ & $0.7$\\
    $\tau_{\psi}$ & $1.0$\\
    $\tau_Q$ & $10.0$ \\
    \hline
   \end{tabular}
   \\
     \begin{tabular}{|c|c|c|}
     \hline
     {\bf Parameter} & 	{small} & {large} \\ 
	\hline 
    $r_{0}$ & $4.0$ & $8.0$\\
    $t_{0}$ & $1.0$ & $1.0$ \\
    $B_{4}$ & $5.0$ & $20.0$\\
    $W$ & $1.0$  & $1.0$\\
	\hline
	\end{tabular}
    \caption{Parameters used in the continuum simulations. The first set of parameters are for generating tactoids. The second set of parameters represent the size of the colloid and the strength of the tactoid-colloid interactions for the two cases shown in Fig.~\ref{fig:tactoidContinuumSimPlots}. There, we show the evolution of tactoid nematic field for two different colloids: a small colloid with weak surface interaction (upper panel) to contrast with a large colloid with strong surface interaction (lower panel). All lengths are in units of grid size, whereas energy and timescales are in arbitrary units. }
\end{table}

The dynamical equations for the density, $\psi$, and the 2D nematic order parameter, ${\bf Q}$, are evolved in time by explicit Euler stepping on a $120 \times 120$ grid with periodic boundary conditions using the pseudospectral method implemented on the numerical package XMDS \cite{XMDS}. The parameters are chosen to be in the nematic-vapor coexistence region of the phase diagram so that tactoids can nucleate, and so that the interface thicknesses are comparable to a few gridpoints. After observing tactoid nucleation and coarsening, a large tactoid is selected and initialized with a colloid at its center. Strong anchoring at the tactoid-colloid interface is enforced with a high value of the parameter $B_4$.

\subsection{Derivation of free energy of a thin splayed nematic layer on the surface of an adhesive colloid} \label{ss:tactoidWettingLayerFEDerivation}
The free energy contributions to the splayed nematic layer of thickness $\lambda$ adhered to the surface of a colloid of radius $a$, as pictured in Fig.~\ref{fig:tactoidTheorySchematic}, will be from the favorable nematic-colloid wetting, the elastic cost due to a splayed nematic, the surface tension cost of the nematic-vapor interface, and the anchoring energy cost due to the deviation of the director field from planar at the nematic-vapor interface. Say that the thin layer covers an area fraction of the colloid $f_l$ which has a value $0$ if none of the surface is covered and $1$ if the entire surface is covered.

Proceeding term-by-term, the adhesive interaction between the colloid and the thin layer will be
\begin{equation}
\begin{split}
F_l^{(w)} &= -w \, f_l \, a^2 \int \textrm{d}S \, 1 \\
 &= -w \, f_l \, 4\pi a^2,
\end{split}
\end{equation}
where the coefficient $w$ is a parameter that controls the strength of the adhesive, or wetting, interaction and where $\textrm{d}S$ is the integral over the solid angle of the surface. The elastic cost of a splayed nematic can be computed
\begin{equation}
\begin{split}
F_l^{(K)} &= K \, f_l \, \int \textrm{d}S \, \int_a^{a+\lambda} \textrm{d} r\, r^2 \left(\nabla \cdot \textbf{n}\right)^2 \\
 &= 4K\lambda \, f_l \, 4\pi ,
 \end{split}
\end{equation}
where $K$ is the Frank elastic constant in the one-constant approximation, and $\textbf{n}$ is the director field of the thin nematic layer, here $\textbf{n} = \hat{\textbf{r}}$. The surface energy due to the nematic-vapor interface for a sphere of radius $a+\lambda$ is simply
\begin{equation}
\begin{split}
F_l^{(\gamma)} &= \gamma \, f_l \left(a + \lambda\right)^2 \int \textrm{d}S \, 1 \\
 &= \gamma \, f_l \, 4\pi \left(a + \lambda\right)^2 ,
\end{split}
\end{equation}
with $\gamma$ the nematic-vapor surface tension. Finally, the nematic-vapor anchoring energy cost is easily computed since we require that $\textbf{n} = \hat{\textbf{r}}$ at the interface:
\begin{equation}
\begin{split}
F_l^{(A_v)} &= \gamma A_v \, f_l \left(a + \lambda\right)^2 \int \textrm{d}S \,  \left(\textbf{v}\cdot\textbf{n}\right)^2 \\
 &=\gamma A_v \, f_l \, 4\pi \left(a + \lambda\right)^2,
\end{split}
\end{equation}
where $\textbf{v} = \hat{\textbf{r}}$ is the surface normal.

We can write the free energy of the thin nematic layer as
\begin{equation}
\begin{split}
F_{l} &= F_l^{(w)} + F_l^{(K)} + F_l^{(\gamma)} + F_l^{(A_v)} \\
 &= \left[-w + \frac{4K}{a}\frac{\lambda}{a} + \gamma\left(1 + A_v\right)\left(1+\frac{\lambda}{a}\right)^2\right] f_{l} \, 4\pi a^2 \\
 &= \left[-w + \epsilon\right] f_{l} \, 4\pi a^2.
\end{split}
\end{equation}

\begin{figure}[h]
\centering
\includegraphics[scale=0.75]{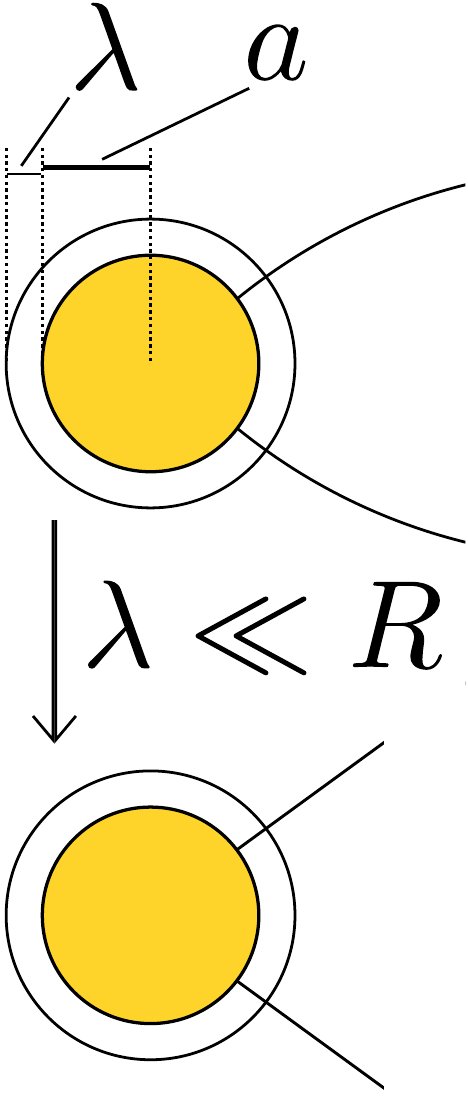}
\caption{When the thickness of the thin splayed nematic layer, $\lambda$ is small relative to the radius of curvature of the associated tactoid, $R$, the curvature can be neglected and the intersection between the tactoid and the nematic layer can be treated as a portion of a cone. The requirement that the tactoid director lines be radial at the colloid surface, discussed in Ref.~\onlinecite{Weirich19}, then sets the ``cone tactoid'' radial at both colloid and splayed nematic layer surfaces and means that the tactoid covers an equal area fraction of both surfaces.}
\label{fig:tactoidColloidWLEquivAreaFraction}
\end{figure}

If a tactoid associates with the colloid-layer system where the entire colloid is covered by the thin nematic layer, the free energy benefit of association now comes from the reduction of the area fraction of the energy penalty $\epsilon f_{l} \, 4\pi a^2$. This can be seen from a few considerations. First, consider the schematic in Fig.~\ref{fig:tactoidColloidWLEquivAreaFraction}, which shows that, for a thin splayed nematic layer, relative to the curvature of the associating tactoid, the area fraction, $f_t$, that the tactoid occupies on the colloid surface and the area fraction the tactoid occupies on the thin splayed nematic layer surface are equivalent. We can think of the tactoid as cutting out a cone shaped region from the nematic layer when it associates with the colloid-layer system. Then, the association of the tactoid will not change the area of the colloid which is wet by a nematic (whether from the tactoid or from the thin splayed nematic layer), so the benefit from the adhesive wetting energy $-w f_l \, 4\pi a^2 = -w (f_l - f_t) \, 4\pi a^2  -w f_t \, 4\pi a^2$ remains constant. On the other hand, the energy penalty due to the wetting layer has its area fraction modified, from its pre-tactoid value of $f_l$ to a post-tactoid value of $f_l - f_t$. Thus, the association of a tactoid leads to an energy benefit of $-\epsilon f_t \, 4\pi a^2$, making $\epsilon$ take the role of an ``effective wetting''.

%

\end{document}